\documentclass[aps,prl,twocolumn,groupedaddress]{revtex4}

\usepackage{graphicx}

\begin{document}

\title{Multi-photon transitions between energy levels\\
in a current-biased Josephson tunnel junction}

\author{A. Wallraff}
\email[]{wallraff@physik.uni-erlangen.de}
\affiliation{Physikalisches Institut III, Universit{\"a}t
Erlangen-N{\"u}rnberg, D-91058 Erlangen, Germany}
\author{T. Duty}
\altaffiliation[Current address ]{Department of Microelectronics and
Nanoscience, MC2, Chalmers University of Technology and G{\"o}teborg
University, S-412 96 Gothenburg, Sweden}
\author{A. Lukashenko}
\altaffiliation{Scientific Center of Physical Technologies,
National Academy of Sciences, 61145 Kharkov, Ukraine}
\author{A.~V. Ustinov}
\affiliation{Physikalisches Institut III, Universit{\"a}t
Erlangen-N{\"u}rnberg, D-91058 Erlangen, Germany}

\date{\today}

\begin{abstract}
The escape of a small current-biased Josephson tunnel junction from
the zero voltage state in the presence of weak microwave radiation is
investigated experimentally at low temperatures.  The measurements of
the junction switching current distribution indicate the macroscopic
quantum tunneling of the phase below a cross-over temperature of
$T^{\star} \approx 280 \, \rm{mK}$.  At temperatures below $T^{\star}$
we observe both single-photon and \emph{multi-photon} transitions
between the junction energy levels by applying microwave radiation in
the frequency range between $10 \, \rm{GHz}$ and $38 \, \rm{GHz}$ to
the junction.  These observations reflect the anharmonicity of the
junction potential containing only a small number of levels.
\end{abstract}


\maketitle

\graphicspath{{Documents:Erlangen:Papers:MultiPhoton:figures:}}

At low temperatures and small damping the dynamics of a current-biased
Josephson junction is governed by the macroscopic quantum mechanics of
the superconducting phase-difference across the junction (see
e.g.~Refs.~\cite{Leggett84,Devoret92} and references therein). 
Macroscopic quantum tunneling of the phase \cite{Martinis87}, energy
level quantization \cite{Martinis87,Silvestrini97} and the effect of
dissipation \cite{Cleland88} have been studied in detail
\cite{Devoret92}.  Josephson junctions are solid-state quantum devices
fabricated with integrated circuit technology.  Their parameters can
be adjusted in a wide range and can be well controlled.  Josephson
junction circuits have been proposed \cite{Bocko97,Mooij99a,Makhlin01}
and recently successfully tested
\cite{Nakamura99,Friedman00,vanderWal00,Vion02,Yu02} as qubits in
quantum information processing \cite{Bennett00}.

In this letter we present experimental evidence of multi-photon
transitions between the ground and the first excited state in a
current-biased Josephson junction.  The experiments have been
performed below the cross-over temperature \cite{Grabert87}, where the
escape of the junction from a metastable state is dominated by quantum
tunneling from the quantized energy levels.  Using a high resolution
measurement \cite{Wallraff02b} of the switching current
\cite{Fulton74}, we detect the multi-photon absorption by monitoring
the decay of the junction from the zero-voltage to the finite voltage
state.

\begin{figure}[tb]
 	\centering
	\includegraphics[width = 1.0\columnwidth]{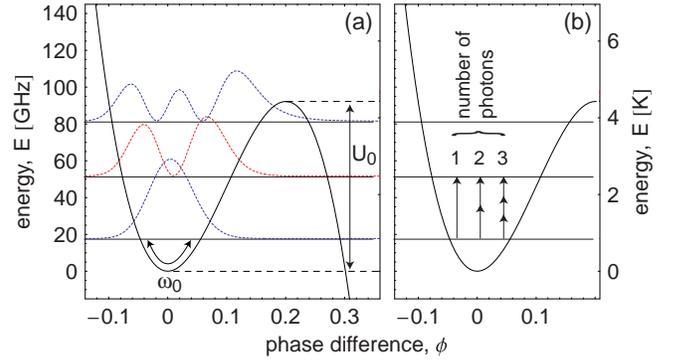}
	\caption{a) The Josephson junction energy $U^{\phi}(\phi)$ calculated
	for the experimental parameters at the bias $\gamma = 0.995$ (solid
	line).  Numerically calculated energy levels (dotted lines) and the
	squared wave functions (dashed lines) are shown.  b) Multi-photon
	transitions between the ground state and the first excited state.}
	\label{fig:potentialQmG995}
\end{figure}

In the Stewart-McCumber model \cite{Stewart68,McCumber68}, the current
biased small Josephson junction is modeled as a particle of mass
$m_{\phi}$ moving in an external washboard potential $U^{\phi} = - E_{J}
(\gamma \phi + \cos \phi)$, see Fig.~\ref{fig:potentialQmG995},
according to the equation of motion
$ m_{\phi} \ddot{\phi} + m_{\phi}({R C})^{-1} \dot{\phi} + \partial
U^{\phi}/\partial \phi = 0.$
Here, the phase difference $\phi$ across the junction represents the
position of the particle.  The particle mass is given by $m_{\phi} = C
(\Phi_{0}/2\pi)^{2}$, where $C$ is the junction capacitance and
$\Phi_{0}$ is the flux quantum.  $E_{J} = \Phi_{0}I_{c}/2\pi$ is the
Josephson coupling energy with the critical current of the junction
$I_{c}$ determining the depth of the potential.  The applied bias
current $I$ normalized as $\gamma = I / I_{c}$ determines the tilt of
the potential and the junction resistance $R$ causes the damping
proportional to the coefficient $1/RC$.

In the absence of thermal or quantum fluctuations and for $\gamma <
1$, the junction is in the zero voltage state, corresponding to the
particle being localized in the the potential well.  At finite
temperatures $T > 0$, the particle may escape from the well at bias
currents $\gamma < 1$ by thermally activated processes
\cite{Ambegaokar69,Fulton74} or by quantum tunneling through the
barrier \cite{Martinis87}.  The rate at which both processes occur
depends on the barrier height $U^{\phi}_{0} = 2 E_{J}
\left[\sqrt{1-\gamma^2} - \gamma \arccos(\gamma) \right] \approx E_{J}
\, {4\sqrt{2}}/{3} \, (1-\gamma)^{3/2}$, the oscillation frequency of
the particle at the bottom of the well $\omega^{\phi}_{0} =
\sqrt{U''^{\phi}(\phi_{0})/m_{\phi}} = \omega_{p}
\left(1-\gamma^{2}\right)^{1/4}$, see Fig.~\ref{fig:potentialQmG995},
and the damping in the junction.  Here $\omega_{p} = \sqrt{2 \pi I_{c}
/ \Phi_{0} C}$ is the Josephson plasma frequency.  At temperatures
below the cross-over temperature $T^{\star}$ \cite{Grabert87} the
quantum tunneling rate dominates the thermal activation rate.  The
quantization of the energy of oscillations of the phase at the bottom
of the well, see Fig.~\ref{fig:potentialQmG995}a, has been observed
both below \cite{Martinis87} and above $T^{\star}$
\cite{Silvestrini97}.

The experiments presented here were performed using a high quality $5
\times 5 \, \rm{\mu m^2}$ tunnel junction fabricated on an oxidized
silicon waver using a standard Nb/Al-AlO$_{x}$/Nb trilayer process. 
The junction had a critical current density of $j_{c} \approx 1.1 \,
\rm{kA/cm^2}$, a capacitance of $C \approx 1.6 \, \rm{pF}$ and a
subgap resistance of $R > 500 \, \rm{\Omega}$ at $T < 2.0 \, \rm{K}$. 
For these sample parameters the expected energy level separation is
larger than $100 \, \rm{GHz}$ at zero bias.  The level width is small
relative to the level spacing, due to the small damping.  The
predicted cross-over temperature $T^{\star}$ is larger than $250 \,
\rm{mK}$.

The sample was mounted in an rf-tight sample box on the cold finger of
a dilution refrigerator.  The dc-bias leads were filtered with
$\pi$-type feedthrough filters at room temperature, RC-filters at the
1 K-pot of the refrigerator and thermocoax filters at the sample box
in order to reduce external electromagnetic interference.  A microwave
signal was fed into the sample cell via a superconducting semi-rigid
coaxial cable.  To reduce the relative level of noise in the microwave
signal, several stages of cold attenuators of a total of $- 40 \,
\rm{dB}$ were used.  We have verified that the power of all harmonics
and subharmonics was at least $80 \, \rm{dB}$ below the fundamental
frequency power.  For the switching current measurements
\cite{Fulton74}, the current was ramped up at a constant rate of
$\dot{I} = 0.245 \, \rm{A/s}$ with a repetition rate of $500 \,
\rm{Hz}$.  The switching current was determined by a measurement of
the time delay between the zero-crossing of the bias current and the
appearance of a voltage across the junction \cite{Wallraff02b}.

The switching current distribution $P(I)$ of the sample in the absence
of microwaves was measured in the temperature range between $4.2 \,
\rm{K}$ and $25 \, \rm{mK}$.  These measurements indicate the thermal
activation of the phase at high temperatures followed by a cross-over
to quantum tunneling around $T^{\star} \approx 280 \, \rm{mK}$
\cite{Wallraff02b}.  In this letter we present measurements performed
at $T < T^{\star}$.

\begin{figure*}[tb]
 	\centering
	\includegraphics[width = \textwidth]{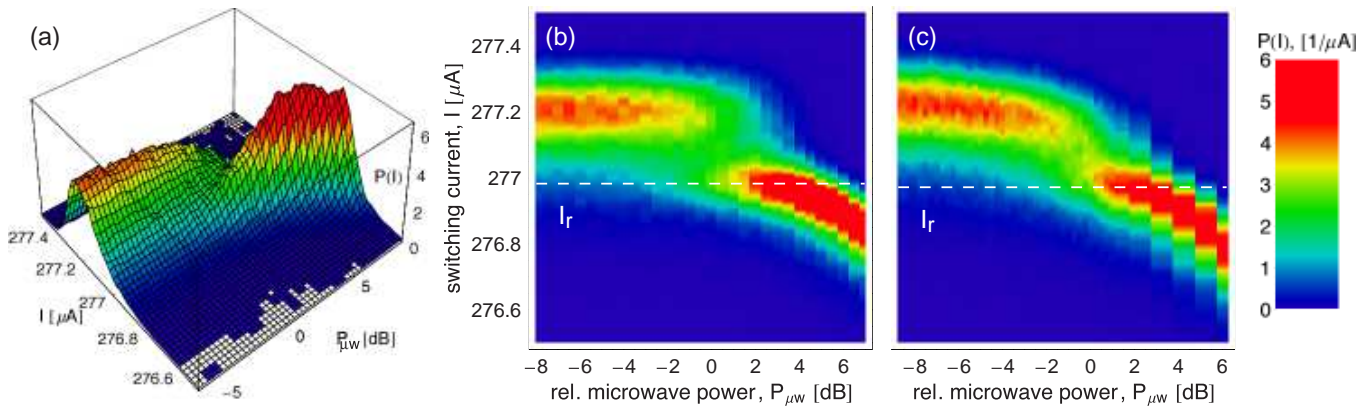}
	\caption{a) 3D plot and b) density plot of the measured $P(I)$
	distribution versus the applied microwave power $P_{\rm{\mu w}}$ at
	$\nu = 36.554 \, \rm{GHz}$ and $T = 100 \, \rm{mK}$.  c) Experimental
	data at $18.399 \, \rm{GHz}$ for the same temperature.  The switching
	probability $P(I)$ is color coded as indicated by the scale.
	\label{fig:SummaryDensSmall}}
\end{figure*}

Microwaves in the frequency range between $10 \, \rm{GHz}$ and $38 \,
\rm{GHz}$ were applied to the sample.  While monitoring the $P(I)$
distributions of the junction, the microwave power $P_{\rm{\mu w}}$
was swept from low values, at which the $P(I)$ distribution is not
changed by the microwaves, to higher values for each chosen frequency. 
At negligibly small microwave powers the $P(I)$ distribution is
essentially determined by the unperturbed quantum tunneling of the
phase from the ground state of the well.  If the microwave power is
increased to substantially populate the excited level, the $P(I)$
distribution becomes double-peaked.  This double-peak structure
smoothly varies with $P_{\rm{\mu w}}$, as shown for $\nu = 36.554 \,
\rm{GHz}$ in Fig.~\ref{fig:SummaryDensSmall}a.  Further increasing the
power, only the pronounced resonant peak is visible in the
distribution.  At this level of power the populations of the ground
and the first excited state are equal, but the tunneling rate from the
excited state is exponentially larger than that from the ground state. 
Thus, the $P(I)$ distribution is dominated by the resonant peak due to
tunneling from the first excited state and the initial peak in the
distribution disappears, see Fig.~\ref{fig:SummaryDensSmall}.  Due to
the resonance excitation of transitions between the two levels, the
switching current distribution at this level of power is \emph{more
narrow} than in the absence of microwaves.  This fact proves that the
measured $P(I)$ distribution in absence of microwaves is not limited
by noise in our experimental setup and that below $T^{\star}$ the
escape indeed occurs due to quantum tunneling through the barrier.

The bias current at which the resonant peak in the $P(I)$ distribution
appears depends strongly on the microwave frequency.  Most strikingly,
we observe resonant peaks at similar or the same bias current for very
different microwave frequencies.  In Figs.~\ref{fig:SummaryDensSmall}b
and c, two representative density plots of the switching current
distributions versus the applied microwave power are shown for the
microwave frequencies $36.554 \, \rm{GHz}$ and $18.399 \, \rm{GHz}$. 
For both frequencies the resonant peaks appear at almost identical
bias currents.  Both sets of data show the \emph{pronounced
narrowing of the distribution} at the resonance.

\begin{figure}[b]
 	\centering
	\includegraphics[width = 0.85 
	\columnwidth]{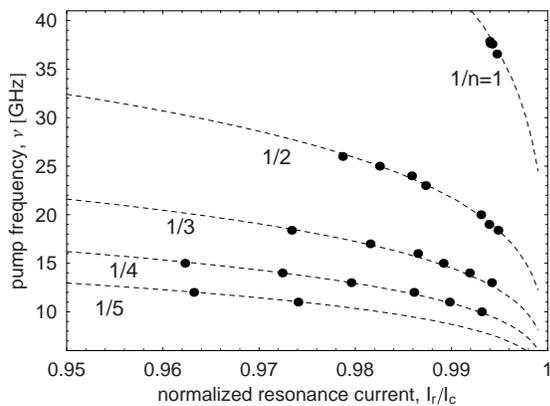}
	\caption{Applied microwave frequency $\nu$ versus normalized resonant bias
	current $I_{\rm{r}}/I_{c}$ (dots).  Dashed lines are a fit of the data
	to $\nu = (1/n) \, \nu_{p} (1-(I_{r}/I_{c})^{2})^{1/4}$. }
	\label{fig:MultiPhotonSumFit}
\end{figure}

\begin{figure}[b]
 	\centering
	\includegraphics[width = 1.0 \columnwidth]{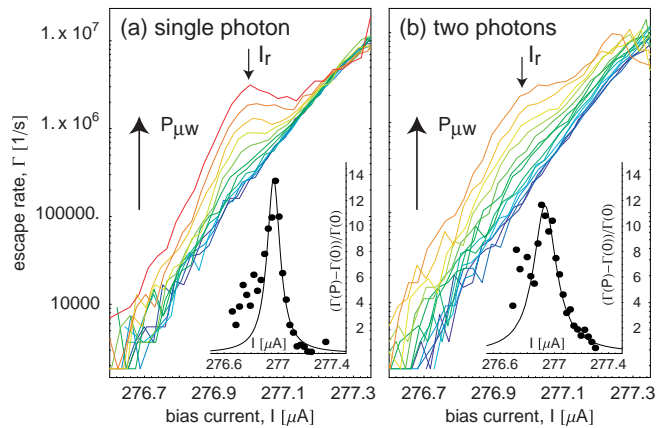}
	\caption{Experimental escape rate $\Gamma(I)$ for (a) single-photon
	and (b) two-photon absorption.  Different curves correspond to
	$P_{\rm{\mu w}}$ being increased (see arrow) from zero to a value at
	which the maximum enhancement $\Gamma(P_{\rm{\mu
	w}}-\Gamma(0))/\Gamma(0)$ is approximately $10$.  The resonance
	current $I_{r}$ is indicated by an arrow.  The insets show the
	enhancement of the escape rate $(\Gamma(P_{\rm{\mu
	w}})-\Gamma(0))/\Gamma(0)$ at the largest displayed value of
	$P_{\rm{\mu w}}$.  Symbols are data, solid lines are fits to a
	Lorentzian line shape.}
	\label{fig:PhotonRate}
\end{figure}

The resonant bias currents $I_{r}$, defined as the current at which
the escape of the phase is maximally enhanced by the microwaves as
indicated in Fig.~\ref{fig:SummaryDensSmall} and
Fig.~\ref{fig:PhotonRate}, were extracted for all measured microwave
frequencies $\nu$, see Fig.~\ref{fig:MultiPhotonSumFit}.  It is
clearly observed that the resonances fall into different groups as
indicated by the dashed lines.  We find that ratios of the resonance
frequencies $\nu$ for a fixed current $I_{r}$ are given with high
accuracy by ratios $m/n$ of two small integer numbers $n$ and $m$,
suggesting that the observed effect is related to multi-photon
transitions between energy levels of the phase.

In parabolic approximation of the potential, one expects the energy level
separation in a small Josephson junction to scale with the applied
bias current as
$
\Delta E = \hbar \omega_{p}
(1-\left(I_{r}/I_{c}\right)^{2})^{1/4}.
$
Therefore resonances with the external applied microwaves are expected
to appear for $n \nu = \Delta E$, where $n$ is the number of photons
absorbed in the transition between two energy levels.  Such
multi-photon transitions between neighboring energy levels are quantum
mechanically allowed \cite{Cohen77} in the anharmonic potential for
the phase of a Josephson junction due to the large diagonal matrix
elements of the excited states.  In Fig.~\ref{fig:MultiPhotonSumFit}
all data are fitted to the single formula $(1/n) \, \nu_{p}
(1-(I_{r}/I_{c})^{2})^{1/4}$, with $\nu_{p} = 116 \, \rm{GHz}$ and
$I_{c} = 278.45 \, \rm{\mu A}$ and $n = 1 \ldots 5$.  The agreement
between the experimental data and this simple formula for the
resonance condition is excellent.  The effective capacitance of the
junction calculated from the fitted zero bias plasma frequency and the
critical current is found to be $C = 1.61 \, \rm{pF}$.  We note that,
the corresponding classical harmonic resonances, arising from the
nonlinearity of the junction, have been observed experimentally
\cite{Dahm68,Bak74} at large applied microwave power and at high
temperatures.

In Fig.~\ref{fig:PhotonRate}, the escape rate $\Gamma(I)$
reconstructed from the $P(I)$ distribution is plotted for a range of
microwave powers for (a) the single- and (b) the two-photon absorption
processes.  In both cases, at low values of $P_{\rm{\mu w}}$ the
escape rate is a monotonic function of the bias current.  With
increasing $P_{\rm{\mu w}}$ a clear resonance develops in the escape
rate.  The resonant current $I_{r}$ is indicated in the plot.

The enhancement $(\Gamma(P_{\rm{\mu w}})-\Gamma(0))/\Gamma(0)$ of the
escape rate in the presence of microwaves of power $P_{\rm{\mu w}}$ is
plotted for both processes in the insets of
Figs.~\ref{fig:PhotonRate}a and b.  $P_{\rm{\mu w}}$ was chosen to
result in a maximum enhancement of roughly $10$ in both cases.  The
data is fitted to a Lorentzian line shape \cite{Martinis87} (solid
line in Fig.~\ref{fig:PhotonRate}).  The width $\delta \nu$ of the
single-photon resonance is in good agreement with the quality factor
of the junction $Q \approx \nu / \delta \nu = 380$ determined from
independent measurements \cite{Wallraff02b}.  For the two-photon
process it is observed that the linewidth is approximately a factor of
two larger than for the single photon process.  For both processes it
is observed that the line width of the transition is independent of
the microwave power.  This indicates that it is entirely limited by
the life time of the excited state and that a coherent broadening of
the resonance was not observable in these measurements.  The
enhancement of the escape rate due to the microwave radiation
increases approximately linearly with $P_{\rm{\mu w}}$ for the single
photon process, whereas it increases roughly as $P_{\rm{\mu w}}^{2}$
for the two-photon process.

We have compared our experimental results with the predictions of the
Larkin-Ovchinnikov theory \cite{Larkin86}.  The bias-current dependent
escape rate $\Gamma(I)$ due to tunneling from all possible energy
levels was calculated using a master equation approach, considering
the occupation of the energy levels in the presence of microwaves at
$T = 100 \, \rm{mK}$.  The energy levels and matrix elements were
determined using the approximations introduced in
Refs.~\cite{Larkin86,Kopietz88,Chow88}, which we found to be
consistent with our direct numerical solutions of the Schr\"odinger
equation for this problem.  For single-photon processes the
microwave-induced transition rates between nearest-neighbor levels $j$
are given by \cite{Larkin86,Kopietz88,Chow88}
\begin{equation}
W_{j,j+1}^{\rm{\mu w}}\propto P_{\rm{\mu w}} \Gamma_j
\left[(2 \pi \nu-\hbar^{-1} \Delta E_{j,j+1})^2+\Gamma_j^2/4\right]^{-1},
\end{equation}
where $\Gamma_j$ is the inverse ``lifetime'' of the $j+1 \rightarrow
j$ transition.

\begin{figure}[t]
 	\centering
	\includegraphics[width = 0.9 \columnwidth]{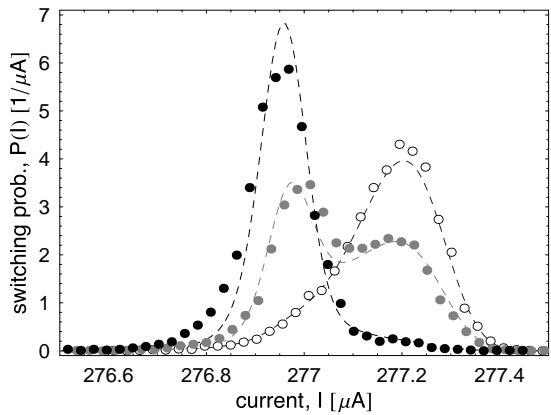}
	\caption{$P(I)$ distributions in the presence of microwave radiation
	at $\nu = 36.554 \, \rm{GHz}$ for $P_{\rm{\mu w}} = -7.5 \, 
	\rm{dB}$ (open points), $1 \, \rm{dB}$ (gray points) and $4 
	\, \rm{dB}$ (black points) measured at $100 \, \rm{mK}$. Dashed curves 
	are calculated according to Larkin-Ovchinnikov theory.}
	\label{fig:Pow36554GHzCutsFit}
\end{figure}

In Fig.~\ref{fig:Pow36554GHzCutsFit}, the measured switching current
distributions for the single photon absorption are fitted to the
calculated distributions at different microwave powers for a junction
capacitance of $C = 1.27 \, \rm{pF}$, a critical current of $I_{c} =
278.25 \, \rm{\mu A}$ and an effective resistance of $R = 500 \,
\rm{\Omega}$ The capacitance obtained in this fit is about 20 \% lower
than that obtained from the plasma frequency.  We suppose this
discrepancy to be due to a microwave-induced level shift which is to
be expected due to the large diagonal matrix elements of the excited
states.  The theory well explains the power dependence of $P(I)$,
strengthening the claim that the resonances are due to the microwave
induced transition of the phase from the ground state to the first
excited state in the well.

In the presented experiments we have found evidence for the
microwave-induced multi-photon transitions between quantized energy
levels of the phase in a current-biased Josephson junction.  This
process could be yet another possible source of decoherence in
microwave driven superconducting qubits.  It could also be used for
manipulating the quantum state of a qubit.

\begin{acknowledgments}
    We are grateful for fruitful discussions with B.~Birnir, M.~Devoret,
    M.~Fistul, S.~Han, J.~Mooij, G.~Sch{\"o}n and S.~Shnirman.  We
    acknowledge the partial financial support by the Deutsche
    Forschungsgemeinschaft (DFG) and by D-Wave Systems Inc.
\end{acknowledgments}


\end{document}